\documentclass[runningheads]{llncs}
\usepackage{cite}
\usepackage{tabularx}
\usepackage{graphicx}
\begin{document}
\title{Deep Transfer Learning For \\ Whole-Brain fMRI Analyses}
\author{Armin W. Thomas\inst{1,2}\orcidID{0000-0002-9947-5705} \and \\ 
Klaus-Robert M{\"u}ller\inst{1,3,4,*} \and
Wojciech Samek\inst{5,*}}
\authorrunning{A. W. Thomas et al.}
\institute{Technische Universit{\"a}t Berlin, 10587 Berlin, Germany \and
Max Planck School of Cognition, 04103 Leipzig, Germany \and
Korea University, 136-713 Seoul, South Korea \and
Max Planck Institute for Informatics, 66123 Saarbr{\"u}cken, Germany \and
Fraunhofer Heinrich Hertz Institute, 10587 Berlin, Germany \\
{*}correspondence to: klaus-robert.mueller@tu-berlin.de, wojciech.samek@hhi.fraunhofer.de}
\maketitle              % typeset the header of the contribution
\begin{abstract}
The application of deep learning (DL) models to the decoding of cognitive states from whole-brain functional Magnetic Resonance Imaging (fMRI) data is often hindered by the small sample size and high dimensionality of these datasets. Especially, in clinical settings, where patient data are scarce. In this work, we demonstrate that transfer learning represents a solution to this problem. Particularly, we show that a DL model, which has been previously trained on a large openly available fMRI dataset of the Human Connectome Project, outperforms a model variant with the same architecture, but which is trained from scratch, when both are applied to the data of a new, unrelated fMRI task. Even further, the pre-trained DL model variant is already able to correctly decode 67.51\% of the cognitive states from a test dataset with 100 individuals, when fine-tuned on a dataset of the size of only three subjects.

\end{abstract}
\section{Introduction}

Over the recent years, deep learning (DL) methods have been shown to outperform more conventional machine learning techniques in a variety of decoding tasks (for a review, see \cite{lecun2015deep}). The success of DL methods is often attributed to their ability to autonomously learn highly abstracted representations of the raw input data, through a hierarchical sequence of non-linear transforms.

While researchers have started exploring the application of DL methods to the analysis of functional Magnetic Resonance Imaging (fMRI) data (e.g., \cite{plis2014deep}), their application to whole-brain fMRI data is still limited (e.g., \cite{huang2017modeling, jang2017task}). Mainly, due to the small sample sizes and high dimensionality of fMRI datasets (and a lack of interpretability of DL models \cite{LapNCOMM19}). Particularly, in clinical settings, where fMRI datasets often only contain 10 - 20 patients and several hundred fMRI samples (i.e., volumes) per patient. Yet, each fMRI volume can easily contain several hundred thousand dimensions (i.e., voxels). In such classification settings, in which the number of data dimensions far exceeds the number of data samples, DL methods, as well as traditional machine learning approaches, are prone to overfitting (for a review, see \cite{lemm2011introduction}).

This problem has been similarly encountered in other research domains (e.g., \cite{oquab2014learning}). Here, researchers have discovered that the successful application of DL models to small datasets can strongly benefit from {\it transfer learning}. Transfer learning describes a process in which a model is trained on one dataset and subsequently applied to another dataset \cite{oquab2014learning}. Thereby, the knowledge about the first dataset, contained in the parameter estimates of the trained model, is utilized to benefit the application of the model to the second dataset. This procedure often drastically improves the classification performance of the model, while also reducing the amount of time and data required to train it.

In this work, we explore whether transfer learning is similarly beneficial for the application of DL models to the decoding of cognitive states (e.g., viewing the image of a face vs the image of a house) from fMRI data. In particular, we show that a DL model that has been trained on the data of six out of seven task-fMRI datasets of the Human Connectome Project database \cite{barch2013function} performs better in decoding the cognitive states underlying a seventh, unrelated task, when compared to a model variant that is trained entirely from scratch on the data of this task. For this comparison, we utilize the DeepLight framework \cite{thomas2018interpretable}, which decodes a cognitive state from whole-brain fMRI data, by combining convolutional and recurrent DL elements (see Fig.\ \ref{fig1} and Section \ref{sec_deeplight}).

\section{Methods}
\subsection{Data}

\subsubsection{Experiment tasks}
\label{sec_tasks}
We analyzed the fMRI data of 400 unrelated participants in the following seven experiment tasks (for further details, see Table 1 and  \cite{barch2013function}):

\begin{itemize}
    \item \textbf{Working Memory (WM):} Participants are asked to decide whether a currently presented image (of body parts, faces, places or tools) is the same as a previously presented target image.
    \item \textbf{Gambling:} Participants are asked to guess whether the value of a card (with values between 1-9) is below or above 5. Participants win or loose if they guess correctly/incorrectly. Trials are neutral if the value of the card is 5.
    \item \textbf{Motor:} Participants are presented with visual cues asking them to tap their left or right fingers, squeeze their left or right toes, or move their tongue.
    \item \textbf{Language:} Participants either hear a brief fable (story trials) or an arithmetic problem (math trials) and are subsequently given a two-alternative forced choice question about the story / arithmetic problem.
    \item \textbf{Social:} Participants are presented with short video clips of objects that either interact in some way or move randomly. Subsequently, participants are asked to decide whether the objects interacted with one another, did not have an interaction, or if they are not sure.
    \item \textbf{Relational:} Participants are presented with different shapes, filled with different textures. In relational trials, participants see a pair of objects at the top of the screen and a pair at the bottom. They are then asked to decide whether the bottom pair differs along the same dimension (shape or texture) as the top pair. In match trials, participants see one object at the top and bottom and are asked to decide whether the objects match on a specified dimension.
    \item \textbf{Emotion:} Participants are asked to decide which of two faces presented on the bottom of the screen matches the face at the top of the screen. The faces have an either angry or fearful expression.
\end{itemize}

\begin{table}
\centering
\caption{Overview of the fMRI Data. For each experiment task, the decoding targets (i.e., cognitive states), the number of decoding targets, as well as the duration of the fMRI data that are included in the analysis are presented.}\label{tab1}
\begin{tabular}{|l|l|l|l|}
\hline
Task &  Decoding targets & Target count & Duration (s)\\
\hline
WM & body, face, place, tool & 4 & 400\\
Gambling & win, loss, neutral & 3 & 224\\
Motor & left/right finger, left/right toe, tongue & 5 & 312\\
Language & story, math & 2 & 480\\
Social & interaction, no interaction & 2 & 200\\
Relational & relational, matching & 2 & 216\\
Emotion & fear, neutral & 2 & 252\\
\hline 
\textbf{Total} & & \textbf{20} & \textbf{2,084}\\
\hline 
\end{tabular}
\end{table}

\subsubsection{FMRI data}
All analyzed fMRI data were provided in a preprocessed format by the Human Connectome Project (HCP), WU Minn Consortium (Principal Investigators: David VanEssen and Kamil Ugurbil; 1U54MH091657) funded by the 16 NIH Institutes and Centers that support the NIH Blueprint for Neuroscience Research; and by the McDonnell Center for Systems Neuroscience at Washington University. Whole-brain EPI acquisitions were acquired with a 32 channel head coil on a modified 3T Siemens Skyra with TR=720 ms and TE=33.1 ms (for further details on fMRI acquisition, see \cite{uugurbil2013pushing}).

\subsubsection{FMRI data preprocessing}
The HCP preprocessing pipeline for fMRI data \cite{glasser2013minimal} includes the following steps: gradient unwarping, motion correction, fieldmap-based EPI distortion correction, brain-boundary based registration of EPI to structural T1-weighted scan, non-linear registration into MNI152 space, and grand-mean intensity normalization (for further details, see \cite{uugurbil2013pushing, glasser2013minimal}). In addition, we applied the following preprocessing: volume-based smoothing with a 3mm Gaussian kernel, linear detrending and standardization of the single voxel signal time-series (resulting in a zero-centered voxel time-series with unit variance) and temporal filtering of the single voxel time-series with a butterworth highpass filter and a cutoff of 128s. We further excluded the first two TRs of every fMRI experiment block (for experiment details, see \cite{barch2013function}) from all analyses, as we did not expect any task-related hemodynamic response within this time period. Each fMRI volume contained 91 x 109 x 91 voxels (X x Y x Z).

\subsubsection{Data splitting}
\label{sec_data_split}
We further divided the fMRI data into a distinct pre-training and test dataset, by assigning the data of the working memory task to the test data and all other experiment tasks to the pre-training data.

\subsection{DeepLight}
\label{sec_deeplight}

\begin{figure}[t!]
\includegraphics[width=\textwidth]{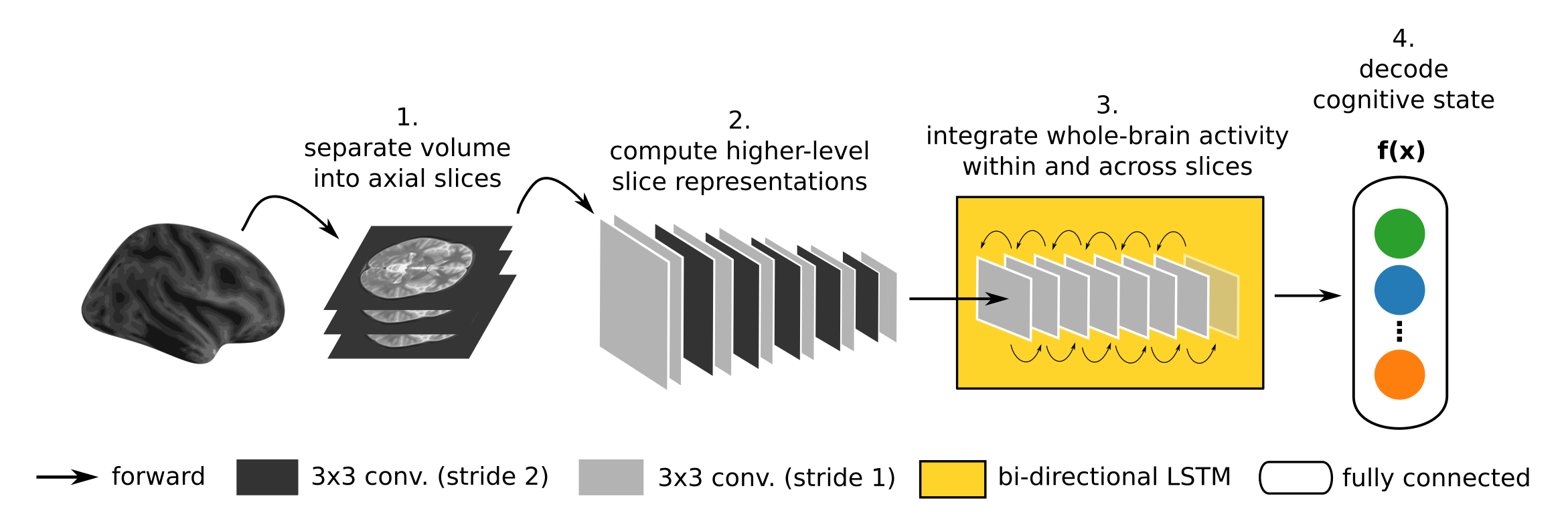}
\caption{Illustration of the DeepLight framework \cite{thomas2018interpretable}. DeepLight first separates a whole-brain fMRI volume into a sequence of axial slices. Each axial slice is then processed by a convolutional feature extractor. The resulting sequence of higher-level axial slice representations is processed by a bi-directional LSTM unit, before a decoding prediction is made through a fully connected softmax output layer.} \label{fig1}
\end{figure}

DeepLight \cite{thomas2018interpretable} consists of three distinct computational modules (see Fig.\ \ref{fig1}). Namely, a feature extractor, an LSTM unit and output layer. To decode a cognitive state, DeepLight first separates a whole-brain fMRI volume into a sequence of axial slices. These slices are then sequentially processed by a convoltional feature extractor. The feature extractor used here consists of the following 12 convolution layers \cite{lecun1995convolutional}: conv3-16(1), conv3-16(1), conv3-16(2), conv3-16(1), conv3-32(2), conv3-32(1), conv3-32(2). conv3-32(1), conv3-64(2), conv3-64(1), conv3-64(2), conv3-64(1) (notation: conv(kernel size) - (number of kernels)(stride size). All convolution kernels were activated through a rectified linear unit function. This sequence of convolution layers resulted in a 768-dimensional representation of each axial volume slice. To integrate the information provided by the resulting sequence of higher-level slice representations into a higher-level representation of the observed whole-brain activity, DeepLight applies a bi-directional LSTM \cite{hochreiter1997long}, containing two independent LSTM units. Each of the two LSTM units contains 64 neurons and iterates through the entire sequence of input slices, but in reverse order (one from bottom-to-top and the other from top-to-bottom). Lastly, to make a decoding decision, DeepLight applies a fully-connected softmax output layer, containing one output neuron per decoding target in the data. 

\subsubsection{Training}
\label{sec_train}
All DeepLight variants that were used in this study were trained as follows (if not reported otherwise): We iteratively trained DeepLight through backpropagation, by the use of the ADAM optimization algorithm, as implemented in tensorflow 1.13. During parameter estimation, we applied dropout regularization to all network layers as follows: We set the dropout probability to 50 \% for the LSTM unit and softmax output layer, For the convolution layers, however, we set the dropout probability to 0\% for the first four convolution layers, 20\% for the next four convolution layers, and 40\% for the last four convolution layers (in line with \cite{thomas2018interpretable}). We further used a learning rate of $1e^{-4}$ and a batch size of 24 fMRI volumes. DeepLight's weights were initialized by the use of a normal-distributed random initialization scheme \cite{glorot2010understanding}.

\section{Results}

\subsection{Pre-training data}
\label{sec_pretrain}

\begin{figure}[t!]
\centering
\includegraphics[width=\textwidth]{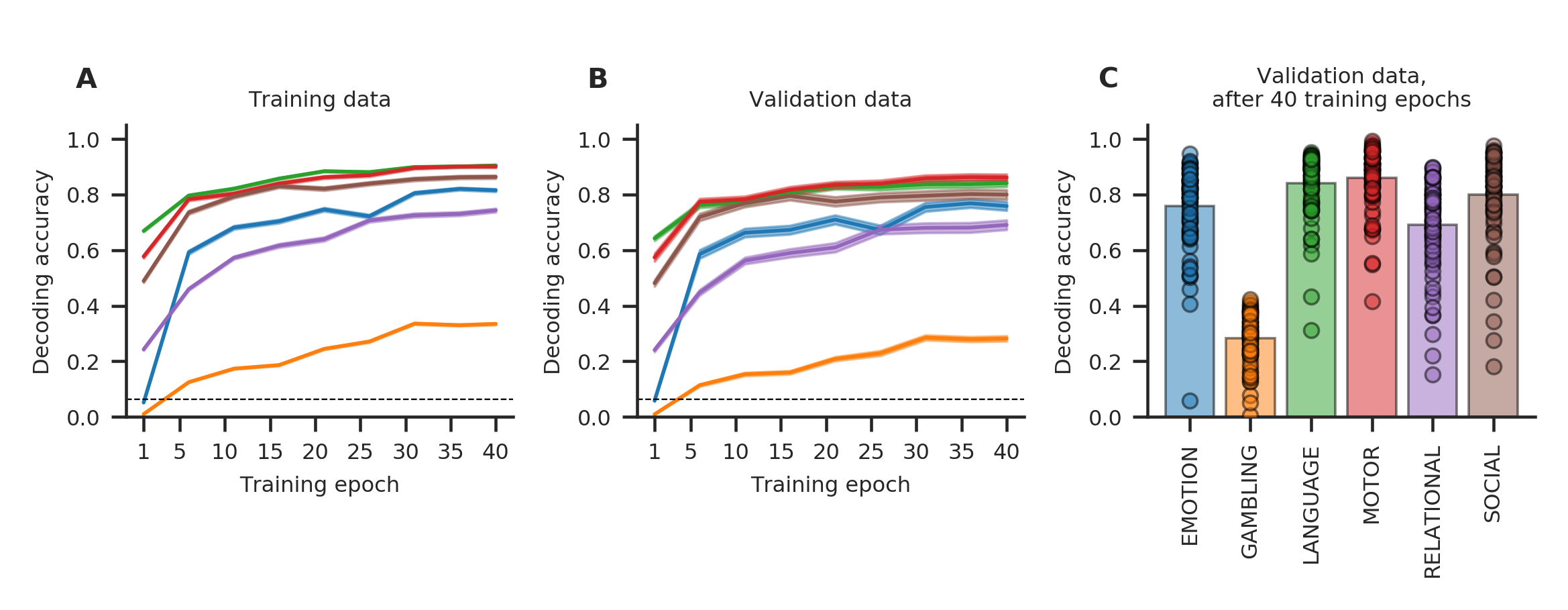}
\caption{DeepLight pre-training statistics. A-B: Mean decoding accuracy in the training (A) and validation (B) data, as a function of training epochs. C: Mean decoding accuracy in the validation data after 40 training epochs. Lines represent grand means, surrounded by standard error bands. Bar heights indicate grand means, while scatter points indicate subject means. Colors indicate tasks. Dashed lines indicate chance level.}
\label{fig2}
\end{figure}

The goal of the first analysis was to pre-train DeepLight on the data of the six tasks contained in the pre-training dataset (see Section \ref{sec_data_split}). To this end, we divided the data within each task into a distinct training and validation dataset, by assigning the data of 300 randomly selected subjects to the training data and the data of the remaining 100 subjects to the validation data. During pre-training, DeepLight's output layer contained 16 neurons, one for each cognitive state of each task in the pre-training dataset (for an overview, see Table 1). Thereby, DeepLight has no knowledge of the underlying tasks and is able to identify an individual's cognitive state without knowing which task the individual performed. Overall, we trained DeepLight for a period of 40 epochs (Fig.\ \ref{fig2}). Each epoch was defined as an iteration over the entire training data.

After 40 training epochs, DeepLight achieved an average decoding accuracy of 76.04\% in the training dataset (Fig.\ \ref{fig2} A) and a decoding accuracy of 70.55\% in the left-out validation data (Fig.\ \ref{fig2} B, C). Interestingly, DeepLight's average decoding accuracy was between 70.00-86.12\% for five out of the six tasks in the validation data, while the decoding accuracy for the sixth task (the gambling task) was only 28.18\%. When excluding the data of the gambling task from the decoding analysis, DeepLight's average decoding accuracy increased to 84.56\% in the training dataset and 79.02\% in the validation data.

\subsection{Test data}

\begin{figure}[t!]
\centering
\includegraphics[width=\textwidth]{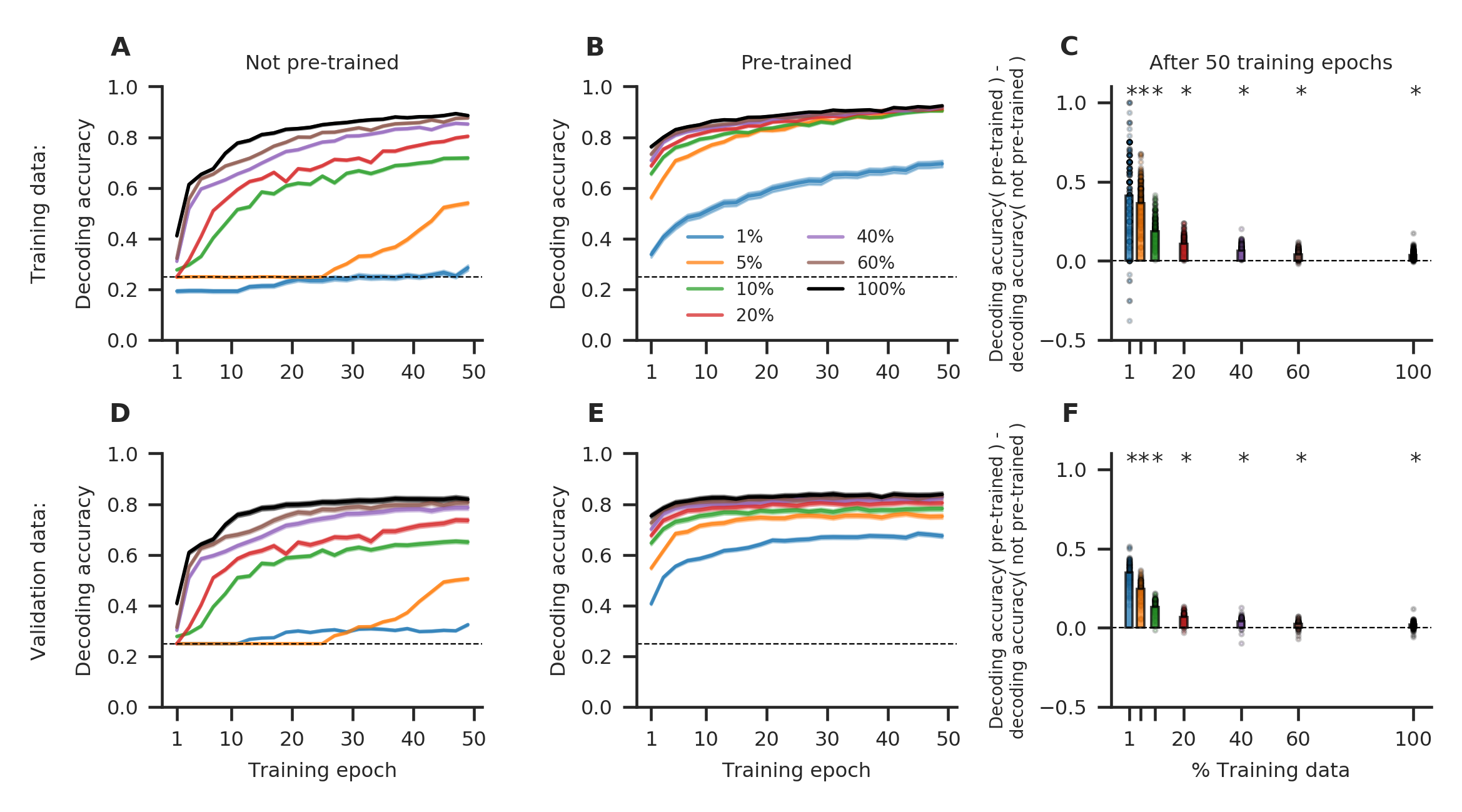}
\caption{Comparison of a "pre-trained" DeepLight variant with a "not pre-trained" variant that is trained entirely from scratch, when both are applied to subsets of 1\%, 5\%, 10\%, 20\%, 40\%, 60\%, and 100\% of the full training dataset ($N=300$) of the test task (the working memory task). A, B, D, E: Decoding accuracy as a function of training epochs in the training (A-B) and validation data (D-E). C, F: Difference in decoding accuracy between the pre-trained and not pre-trained DeepLight variant after 50 training epochs. Stars indicate a statistically meaningful difference in a t-test using Bonferroni adjusted alpha levels of 0.05/7. Colors indicate the fraction of training data that is used. Lines show grand means with standard error bands surrounding them. Bar heights indicate grand means. Scatter points indicate  subject means.}
\label{fig3}
\end{figure}

The goal of the second analysis was to explore the benefits of transfer learning for the application of DL models to fMRI data. To this end, we compared the performance of the pre-trained DeepLight variant (see Section \ref{sec_pretrain}) to that of a variant that was trained entirely from scratch, when both are applied to the data of the left-out test task (the working memory task, see Section \ref{sec_data_split}). We again divided the data of the test task into a separate training and validation dataset, by randomly assigning 300 distinct subjects to the training data and the remaining 100 to the validation data. We then trained a DeepLight variant with transfer learning, and one without, on the training data of the test task. The output layer of both variants was set to contain four neurons (one per decoding target in the working memory task, see Table 1). Otherwise, the architecture and training procedures (see Section \ref{sec_train}) of both variants were identical. 

The first variant ("not pre-trained") does not apply transfer learning and was trained entirely from scratch, with weights initialized according to the normal-distributed random initialization scheme \cite{glorot2010understanding}. After 50 training epochs, this variant achieved an average decoding accuracy of 88.57\% in the training data of the test task (Fig.\ \ref{fig3} A) and 81.91\% in the validation data (Fig.\ \ref{fig3} D). The second variant ("pre-trained") applies transfer learning and is based on the DeepLight variant that we previously trained on the pre-training dataset (see Section \ref{sec_data_split} and \ref{sec_pretrain}). Particularly, we initialized the parameters of all network layers, except for the output layer (Fig.\ \ref{fig1}), to those weights obtained by the pre-trained DeepLight variant and only initialized the weights of the output layer according to the normal-distributed random initialization scheme \cite{glorot2010understanding}. After 50 training epochs, the pre-trained variant achieved an average decoding accuracy of 92.43\% in the training data of the test task (Fig.\ \ref{fig3} B) and 83.83\% in the validation data (Fig.\ \ref{fig3} E) and thereby performed meaningfully better in decoding the cognitive states from the validation data than the not pre-trained variant ($t(99)=8.42, p < 0.0001$), Fig.\ \ref{fig3} F). 

We were further interested in exploring how both DeepLight variants performed, when trained on smaller fractions of the original training dataset of the test task. Therefore, we repeatedly trained both variants on 1\%, 5\%, 10\%, 20\%, 40\% and 60\% of the full training dataset of the test task ($N=300$), while evaluating their performance on the full validation data of the test task ($N=100$). Overall, the pre-trained variant consistently achieved higher decoding accuracies in the training (Fig.\ \ref{fig3} C) and validation (Fig.\ \ref{fig3} F) data, and required less training time, when compared to the not pre-trained variant. Importantly, the pre-trained DeepLight variant already achieved an average decoding accuracy of 67.51\% (Fig.\ \ref{fig3} E) in the validation data, when being trained on only 1\% of the training dataset (equal to the data of three subjects). The not pre-trained variant, on the other hand, achieved a decoding accuracy of only 32.49\% (Fig.\ \ref{fig3} D), when being trained on 1\% of the training data and thereby performed meaningfully worse (the pre-trained DeepLight variant outperformed the not pre-trained variant by 35.02\% ($t(99)=49.68, p<0.0001$)). Lastly, we also tested how much of the training data the pre-trained DeepLight variant requires to performs as well as (or better than) the not pre-trained variant that has been trained on the full training data. Interestingly, the pre-trained variant already achieved a meaningfully better decoding accuracy than the not pre-trained variant (which was trained on the full training dataset), when the pre-trained variant was trained on only 40\% of the training data ($t(99)=2.82, p=0.0057$).

\section{Conclusion}
The broad application of DL models to fMRI data has long been hindered by the small sample size, and high dimensionality, of typical fMRI datasets. Here, we have demonstrated that transfer learning is strongly beneficial for the application of DL models to small fMRI datasets. A DL model that has been pre-trained on a large, openly available fMRI dataset, generally requires less training data and time, and achieves higher decoding accuracies, when compared to a model variant with the same architecture that is trained entirely from scratch. Even further, the pre-trained model variant already performs well in decoding the cognitive states of 100 individuals in an unrelated fMRI task, when fine-tuned on a dataset of the size of only three subjects.
\bibliographystyle{splncs04}
\bibliography{main}
\end{document}